\documentclass[journal,onecolumn,draftcls]{IEEEtran}
\usepackage{ifpdf}
\ifpdf
    \usepackage[pdftex]{graphicx}
    \graphicspath{{figs/}{matlab/}}   
    \usepackage[update]{epstopdf}
\else
	\usepackage{graphicx}
	\graphicspath{{figs/}{matlab/}}   
\fi

\usepackage{array}
\usepackage{amsmath}
\usepackage{amsfonts}
\usepackage{amssymb}
\usepackage{amstext}
\usepackage{latexsym}
\usepackage{color}
\usepackage{cite,url}
\usepackage{setspace}
\newtheorem{thm}{Theorem}
\newtheorem{prop}{Proposition}
\newtheorem{lemma}{Lemma}

\newtheorem{coro}[lemma]{Corollary}

\usepackage{pdflscape}
\allowdisplaybreaks
\usepackage{multirow}

\DeclareMathSizes{10}{9}{8}{7} 

\begin{document}

\title{\huge Multilevel Diversity Coding with Secure Regeneration: Separate Coding Achieves the MBR Point}
\author{Shuo Shao,~\IEEEmembership{Student Member,~IEEE}, Tie Liu,~\IEEEmembership{Senior Member,~IEEE}, \\Chao Tian,~\IEEEmembership{Senior Member,~IEEE}, and Cong Shen,~\IEEEmembership{Senior Member,~IEEE}
\thanks{S.~Shao, T.~Liu, and C.~Tian are with the Department of Electrical and Computer Engineering, Texas A\&M University, College Station, TX 77843, USA (e-mail: \{shaoshuo,tieliu,chao.tian\}@tamu.edu).}
\thanks{C.~Shen is with School of Information Science and Technology, University of Science and Technology of China (USTC), Hefei, China 230027, (e-mail: congshen@ustc.edu.cn).}}

\textfloatsep=0.15cm
\intextsep=0.15cm
\abovecaptionskip=0.15cm
\belowcaptionskip=0.15cm
\setlength{\abovedisplayskip}{5pt}
\setlength{\belowdisplayskip}{5pt}

\maketitle
\begin{abstract}
The problem of multilevel diversity coding with secure regeneration (MDC-SR) is considered, which includes the problems of multilevel diversity coding with regeneration (MDC-R) and secure regenerating code (SRC) as special cases. Two outer bounds are established, showing that separate coding of different messages using the respective SRCs can achieve the minimum-bandwidth-regeneration (MBR) point of the achievable normalized storage-capacity repair-bandwidth tradeoff regions for the general MDC-SR problem. The core of the new converse results is an exchange lemma, which can be established using Han's subset inequality.
\end{abstract}


%


\section{Introduction}\label{sec:Intro}
Diversity coding and node repair are two fundamental ingredients of reliable distributed storage systems. While the study of diversity coding has been in the literature for decades \cite{Sin-IT64,Roc-Thesis92,RYH-IT97,YZ-IT99,MTD-IT10,JML-IT14}, systematic studies of node repair mechanisms were started only recently by Dimakis et al. in their pioneering work \cite{Dimakis-IT10}. A particular model, which was first introduced in \cite{Dimakis-IT10} and has since received a significant amount of attention in the literature \cite{Kumar-IT11,Cadambe-IT13,Tian-JSAC13,Goparaju-ISIT14,Duursma-P14,Prakash-ISIT15,Elyasi-ISIT15,Tian-IT15,Ye-IT17}, is the so-called {\em (exact-repair) regenerating code (RC)} problem. 

More specifically, in an $(n,k,d)$ RC problem, a file $\mathsf{M}$ of size $B$ is to be encoded in a total of $n$ distributed storage nodes, each of capacity $\alpha$. The encoding needs to ensure that the file $\mathsf{M}$ can be perfectly recovered by having full access to any $k$ out of the total $n$ storage nodes. In addition, when a node failure occurs, it is required that the data originally stored in this failed node can be recovered by downloading data of size $\beta$ each from any $d$ remaining nodes. An interesting technical challenge is to characterize the optimal {\em tradeoffs} between the node capacity $\alpha$ and the download bandwidth $\beta$ in satisfying both the file-recovery and node-repair requirements. However, despite intensive research efforts that have yielded many interesting and highly non-trivial partial results including a precise characterization of the {\em minimum-storage-regenerating (MSR)} and the {\em minimum-bandwidth-regenerating (MBR)} rate points \cite{Kumar-IT11,Cadambe-IT13,Tian-JSAC13,Goparaju-ISIT14,Duursma-P14,Prakash-ISIT15,Elyasi-ISIT15,Tian-IT15,Ye-IT17,LiTangTian:ISIT17}, the optimal tradeoffs between the node capacity $\alpha$ and the download bandwidth $\beta$ have {\em not} been fully understood for the general RC problem.

More recently, two extensions of the RC problem, namely {\em multilevel diversity coding with regeneration (MDC-R)} and {\em secure regenerating code (SRC)}, have also been studied in the literature. The problem of MDC-R was first introduced by Tian and Liu \cite{Tian-IT16}. In an $(n,d)$ MDC-R problem, a total of $d$ independent files $\mathsf{M}_1,\ldots,\mathsf{M}_d$ of size $B_1,\ldots,B_d$, respectively, are to be stored in $n$ distributed storage nodes, each of capacity $\alpha$. The encoding needs to ensure that the file $\mathsf{M}_j$ can be perfectly recovered by having full access to any $j$ out of the total $n$ storage nodes for any $j\in\{1,\ldots,d\}$. In addition, when a node failure occurs, it is required that the data originally stored in this failed node can be recovered by downloading data of size $\beta$ each from any  $d$ remaining nodes.

Clearly, an $(n,k,d)$ RC problem can be viewed as an $(n,d)$ MDC-R problem with {\em degenerate} messages $(\mathsf{M}_j: j \neq k)$ (i.e., $B_j=0$ for all $j\neq k$). Therefore, from the code construction perspective, it is natural to consider the so-called {\em separate coding} scheme, i.e., to construct a code for the $(n,d)$ MDC-R problem, we can simply use an $(n,j,d)$ RC to encode the file $\mathsf{M}_j$ for each $j\in\{1,\ldots,d\}$, and the coded messages for each file remain separate when stored in the storage nodes and during the repair processes. However, despite being a natural scheme, it was shown in \cite{Tian-IT16} that separate coding is in general {\em suboptimal} in achieving the optimal tradeoffs between the normalized storage-capacity and repair-bandwidth. On the other hand, it has been shown that separate coding can, in fact, achieve both the MSR \cite{Tian-IT16} and the MBR \cite{Shao-CISS16} points of the achievable normalized storage-capacity and repair-bandwidth tradeoff region for the general MDC-R problem.

The problem of SRC is an extension of the RC problem that further requires security guarantees during the repair processes. More specifically, the $(n,k,d,\ell)$ SRC problem that we consider is the $(n,k,d)$ RC problem \cite{Dimakis-IT10,Kumar-IT11,Cadambe-IT13,Tian-JSAC13,Goparaju-ISIT14,Duursma-P14,Prakash-ISIT15,Elyasi-ISIT15,Tian-IT15,Ye-IT17}, with the additional constraint that the file $\mathsf{M}$ needs to be kept {\em information-theoretically} secure against an eavesdropper, which can access the data downloaded to regenerate a total of $\ell$ different failed nodes under all possible repair groups. Obviously, this is only possible when $\ell<k$. Furthermore, when $\ell=0$, the secrecy requirement degenerates, and the $(n,k,d,\ell)$ SRC problem reduces to the $(n,k,d)$ RC problem {\em without} any repair secrecy requirement.

Under the additional secrecy requirement ($\ell \geq 1$), the optimal tradeoffs between the node capacity $\alpha$ and repair bandwidth $\beta$ have been studied in \cite{Pawar-ISIT10,Pawar-IT11,Shah-Globecom11,Goparaju-NetCod13,Rawat-IT14,Tandon-IT16,Ye-ISIT16,Shao-IT17,Shah-IT17}. In particular, Shah, Rashmi and Kumar \cite{Shah-Globecom11} showed that a particular tradeoff point (referred to as the {\em SRK} point) can be achieved by extending an MBR code based on the product-matrix construction proposed in \cite{Kumar-IT11}. Later, it was shown \cite{Shao-IT17} that for any given $(k,d)$ pair, there is a lower bound on $\ell$, denoted by $\ell^*(k,d)$, such that when $\ell \geq \ell^*(k,d)$, the SRK point is the {\em only} conner point of the tradeoff region for the $(n,k,d,\ell)$ SRC problem. On the other hand, when $1 \leq \ell < \ell^*(k,d)$, it is possible that the tradeoff region features {\em multiple} corner points. However, a precise characterization of the tradeoff region, including both the MSR and the MBR points, remains missing in general.

In this paper, we introduce the problem of {\em multilevel diversity coding with secure regeneration (MDC-SR)}\footnote{The problem of secure multilevel diversity coding {\em without} any node regeneration requirement has been considered in \cite{Balasubramanian-IT13,Jiang-IT14}.}, which includes the problems of MDC-R and SRC as two special cases. Similar to the MDC-R problem, it is natural to consider the separate coding scheme for the MDC-SR problem as well. Our main result of the paper is to show that the optimality of separate coding in terms of achieving the MBR point of the achievable normalized storage-capacity and repair-bandwidth tradeoff region extends more generally from the MDC-R problem to the MDC-SR problem. When specialized to the SRC problem, this shows conclusively that the SRK point \cite{Shah-Globecom11} is, in fact, the MBR point of the achievable normalized storage-capacity and repair-bandwidth tradeoff region, {\em regardless} of the number of corner points of the tradeoff region.

From the technical viewpoint, this is mainly accomplished by establishing two outer bounds (one of them must be ``horizontal", i.e., on the normalized repair-bandwidth {\em only}) on the achievable normalized storage-capacity and repair-bandwidth tradeoff region, which intersect precisely at the superposition of the SRK points. The core of the new converse results is an {\em exchange} lemma, which we establish by exploiting the built-in symmetry of the problem via Han's subset inequality \cite{Han-IC78}. The meaning of ``exchange" will be clear from the statement of the lemma. The lemma {\em only} relies on the functional dependencies for the repair processes and might be useful for solving some other related problems as well.

The rest of the paper is organized as follows. In Section~\ref{sec:PF} we formally introduce the problem of MDC-SR and the separate coding scheme. The main results of the paper are then presented in Section~\ref{sec:Main}. In Section~\ref{sec:Proof}, we introduce the exchange lemma and use it to establish the main results of the paper. Finally, we conclude the paper in Section~\ref{sec:Con}.

{\em Notation}. Sets and random variables will be written in calligraphic and sans-serif fonts respectively, to differentiate from the real numbers written in normal math fonts. For any two integers $t \leq t'$, we shall denote the set of consecutive integers $\{t,t+1,\ldots,t'\}$ by $[t:t']$. The use of the brackets will be surpressed otherwise.

\section{The MDC-SR Problem}\label{sec:PF}
Let $(n,d,N_1,\ldots,N_d,K,T,S)$ be a tuple of positive integers such that $d<n$. Formally, an $(n,d,N_1,\ldots,N_d,K,T,S)$ code consists of:
\begin{itemize}
\item for each $i \in [1:n]$, a {\em message-encoding} function $f_i: \left(\prod_{j=1}^{d}[1:N_j]\right)\times[1:K] \rightarrow [1:T]$;
\vspace{4pt}
\item for each $\mathcal{A} \subseteq [1:n]: |\mathcal{A}|\in[1:d]$, a {\em message-decoding} function $g_\mathcal{A}: [1:T]^{|\mathcal{A}|} \rightarrow [1:N_{|\mathcal{A}|}]$;
\vspace{4pt}
\item for each $\mathcal{B}\subseteq [1:n]: |\mathcal{B}|=d$, $i' \in \mathcal{B}$, and $i \in [1:n]\setminus\mathcal{B}$, a {\em repair-encoding} function $f^{\mathcal{B}}_{i' \rightarrow i}: [1:T] \rightarrow [1:S]$;
\vspace{4pt}
\item for each $\mathcal{B}\subseteq [1:n]: |\mathcal{B}|=d$ and $i \in [1:n]\setminus\mathcal{B}$, a {\em repair-decoding} function $g^{\mathcal{B}}_i:[1:S]^d \rightarrow [1:T]$. 
\end{itemize} 

For each $j\in[1:d]$, let $\mathsf{M}_j$ be a message that is uniformly distributed over $[1:N_j]$. The messages $\mathsf{M}_1,\ldots,\mathsf{M}_d$ are assumed to be mutually independent. Let $\mathsf{K}$ be a random key that is uniformly distributed over $[1:K]$ and independent of the messages $(\mathsf{M}_1,\ldots,\mathsf{M}_d)$. For each $i\in[1:n]$, $\mathsf{W}_i\triangleq f_i(\mathsf{M}_1,\ldots,\mathsf{M}_d,\mathsf{K})$ is the data stored at the $i$th storage node, and for each $\mathcal{B}\subseteq [1:n]: |\mathcal{B}|=d$, $i' \in \mathcal{B}$, and $i \in [1:n]\setminus\mathcal{B}$, $\mathsf{S}^{\mathcal{B}}_{i'\rightarrow i}\triangleq f^{\mathcal{B}}_{i'\rightarrow i}(\mathsf{W}_{i'})$ is the data downloaded from the $i'$th storage node in order to regenerate the data originally stored at the $i$th storage node under the context of repair group $\mathcal{B}$. Obviously, 
\begin{align*}
(B_j&=\log{N_j}: j\in[1:d]), \quad \alpha=\log{T}, \quad \mbox{and} \;\; \beta=\log{S}
\end{align*}
represent the message sizes, storage capacity, and repair bandwidth, respectively. 

\begin{figure*}[t!]
\centering
\includegraphics[width=0.58\linewidth,draft=false]{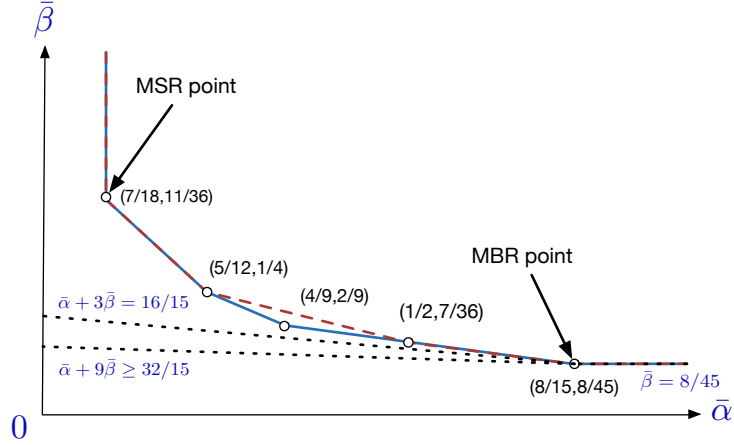}
\caption{The optimal tradeoff curve between the normalized storage-capacity $\bar{\alpha}$ and repair-bandwidth $\bar{\beta}$ (the solid line) and the best possible tradeoffs that can be achieved by separate coding (dashed line) for the $(4,3)$ MDC-R problem with $(\bar{B}_1,\bar{B}_2,\bar{B}_3)=(0,1/3,2/3)$ \cite{Tian-IT16}. The outer bounds \eqref{eq:B1}, \eqref{eq:B2} and \eqref{eq:B7} are evaluated as $\bar{\beta} \geq 8/45$, $\bar{\alpha}+3\bar{\beta} \geq 16/15$, and $\bar{\alpha}+9\bar{\beta} \geq 32/15$, respectively. When set as equalities, they intersect precisely at the MBR point $(8/15,8/45)$.}
\label{fig1}
\end{figure*}

A normalized message-rate storage-capacity repair-bandwidth tuple $(\bar{B}_{\ell+1},\ldots,\bar{B}_d,\bar{\alpha},\bar{\beta})$ is said to be {\em achievable} for the $(n,d,\ell)$ MDC-SR problem if an $(n,d,1,\ldots,1,N_{\ell+1},\ldots,N_d,K,T,S)$ code (i.e., $N_j=1$ for all $j\in[1:\ell]$) can be found such that the following requirements are satisfied:
\begin{itemize}
\item rate normalization
\begin{align}
\frac{\alpha}{\sum_{t=\ell+1}^dB_t}=\bar{\alpha}, \; 
\frac{\beta}{\sum_{t=\ell+1}^dB_t}=\bar{\beta}, \; \frac{B_j}{\sum_{t=\ell+1}^dB_t}=\bar{B}_j\label{eq:Rate}
\end{align}
for any $j\in[\ell+1:d]$; 
\vspace{4pt}
\item message recovery
\begin{align}
\mathsf{M}_{|\mathcal{A}|} =g_\mathcal{A}(\mathsf{W}_i:i \in\mathcal{A})
\label{eq:MessageRecovery}
\end{align}
for any $\mathcal{A} \subseteq [1:n]: |\mathcal{A}|\in[\ell+1:d]$;
\vspace{4pt}
\item node regeneration
\begin{align}
\mathsf{W}_i =g^{\mathcal{B}}_i(\mathsf{S}^{\mathcal{B}}_{i' \rightarrow i}:i'\in \mathcal{B})
\label{eq:NodeRegen}
\end{align}
for any $\mathcal{B}\subseteq [1:n]: |\mathcal{B}|=d$ and $i \in [1:n]\setminus\mathcal{B}$;
\vspace{4pt}
\item repair secrecy 
\begin{align}
I((\mathsf{M}_{\ell+1},\ldots,\mathsf{M}_d);(\mathsf{S}_{\rightarrow i}:i\in \mathcal{E}))=0\label{eq:RepairSecrecy}
\end{align}
for any $\mathcal{E}\subseteq [1:n]$ such that $|\mathcal{E}|=\ell$, where $\mathsf{S}_{\rightarrow i} :=(\mathsf{S}^{\mathcal{B}}_{i'\rightarrow i}:\mathcal{B}\subseteq [1:n], \; |\mathcal{B}|=d, \; \mathcal{B}\not\ni i, \; i' \in \mathcal{B})$ is the collection of data that can be downloaded from the other nodes to regenerate node $i$.
\end{itemize}
The closure of all achievable $(\bar{B}_{\ell+1},\ldots,\bar{B}_d,\bar{\alpha},\bar{\beta})$ tuples is the {\em achievable normalized message-rate storage-capacity repair-bandwidth tradeoff region} $\mathcal{R}_{n,d,\ell}$ for the $(n,d,\ell)$ MDC-SR problem. For a fixed normalized message-rate tuple $(\bar{B}_{\ell+1},\ldots,\bar{B}_d)$, the {\em achievable normalized storage-capacity repair-bandwidth tradeoff region} is the collection of all normalized storage-capacity repair-bandwidth pairs $(\bar{\alpha},\bar{\beta})$ such that $(\bar{B}_{\ell+1},\ldots,\bar{B}_d,\bar{\alpha},\bar{\beta}) \in \mathcal{R}_{n,d,\ell}$ and is denoted by $\mathcal{R}_{n,d,\ell}(\bar{B}_{\ell+1},\ldots,\bar{B}_d)$.

Based on the above problem formulation, it should be clear that the MDC-RC can be specialized to various cases that have been considered in the literature: 
\begin{itemize}
\item[1)] the achievable normalized storage-capacity repair-bandwidth tradeoff region $\mathcal{R}_{n,d}(\bar{B}_1,\ldots,\bar{B}_d)$ of the $(n,d)$ MDC-R problem is simply $\mathcal{R}_{n,d,0}(\bar{B}_1,\ldots,\bar{B}_d)$ for any given normalized message-rate tuple $(\bar{B}_1,\ldots,\bar{B}_d)$; 
\vspace{4pt}
\item[2)] the achievable normalized storage-capacity repair-bandwidth tradeoff region $\mathcal{R}_{n,k,d,\ell}$ of the $(n,k,d,\ell)$ SRC problem is simply $\mathcal{R}_{n,d,\ell}(0,\ldots,0,\bar{B}_k=1,0,\ldots,0)$.
\vspace{4pt}
\item[3)] the achievable normalized storage-capacity repair-bandwidth tradeoff region $\mathcal{R}_{n,k,d}$ of the $(n,k,d)$ RC problem is simply $\mathcal{R}_{n,d}(0,\ldots,0,\bar{B}_k=1,0,\ldots,0)$ or, equivalently, $\mathcal{R}_{n,k,d,0}$.
\end{itemize}

A simple and natural strategy for constructing a code for the $(n,d,\ell)$ MDC-SR problem is to use to an $(n,j,d,\ell)$ SRC to encode the message $\mathsf{M}_j$ separately for each $j\in[\ell+1:d]$. Since the coded data are kept separate during the encoding, decoding and repair processes, we have
\begin{align*}
K=\prod_{j=\ell+1}^{d}K_j, \;\; T=\prod_{j=\ell+1}^{d}T_j, \;\; \mbox{and} \;\; S=\prod_{j=\ell+1}^{d}S_j.
\end{align*}
Thus, for the general MDC-SR problem, the {\em separate coding normalized storage-capacity repair-bandwidth tradeoff region} $\hat{\mathcal{R}}_{n,d,\ell}(\bar{B}_{\ell+1},\ldots,\bar{B}_d)$ for a fixed normalized message-rate tuple $(\bar{B}_{\ell+1},\ldots,\bar{B}_d)$ is given by:
\begin{align}
\left(\left(\sum_{j=\ell+1}^d\bar{\alpha}_j\bar{B}_j,\sum_{j=\ell+1}^d\bar{\beta}_j\bar{B}_j\right):(\bar{\alpha}_j,\bar{\beta}_j)\in\mathcal{R}_{n,j,d,\ell}\right).\label{eq:SC}
\end{align}
As mentioned previously, when $\ell=0$, the repair secrecy requirement \eqref{eq:RepairSecrecy} degenerates, and the $(n,d,\ell)$ MDC-SR problem reduces to the $(n,d)$ MDC-R problem. In this case, it was shown in \cite{Shao-CISS16} that any achievable normalized message-rate storage-capacity repair-bandwidth tuple $(\bar{B}_1,\ldots,\bar{B}_d,\bar{\alpha},\bar{\beta}) \in \mathcal{R}_{n,d}$ must satisfy:
\begin{align}
\bar{\beta} & \geq \sum_{j=1}^{d}T_{d,j}^{-1}\bar{B}_j \label{eq:B1}\\
\mbox{and} \quad \bar{\alpha}+\frac{d(d-1)}{2}\bar{\beta} & \geq \frac{d(d+1)}{2}\sum_{j=1}^{d}T_{d,j}^{-1}\bar{B}_j\label{eq:B2}
\end{align}
where $T_{d,j}:=\sum_{t=1}^{j}(d+1-t)$. When set as equalities, the intersection of \eqref{eq:B1} and \eqref{eq:B2} is given by:
\begin{align*}
\left(\bar{\alpha},\bar{\beta}\right)
&=\left(d\sum_{j=1}^{d}T_{d,j}^{-1}\bar{B}_j,\sum_{j=1}^{d}T_{d,j}^{-1}\bar{B}_j\right).
\end{align*}
For any $j\in [1:d]$, the MBR point for the $(n,j,d)$ RC problem can be written as \cite{Kumar-IT11}
\begin{align}
\left(dT_{d,j}^{-1},T_{d,j}^{-1}\right)\in \mathcal{R}_{n,j,d}.\label{eq:MBR}
\end{align}
We may thus conclude immediately from \eqref{eq:SC} (with $\ell=0$) that separate coding can achieve the MBR point for the general MDC-R problem. 

Fig.~\ref{fig1} shows the optimal tradeoff curve between the normalized storage-capacity and repair-bandwidth and the best possible tradeoffs that can be achieved by separate coding for the $(4,3)$ MDC-R problem with $(\bar{B}_1,\bar{B}_2,\bar{B}_3)=(0,1/3,2/3)$ \cite{Tian-IT16}. Clearly, for this example, separate coding is strictly suboptimal when $\bar{\alpha}\in(5/12,1/2)$. On the other hand, when $\bar{\alpha} \leq 5/12$ or $\bar{\alpha} \geq 1/2$, separate coding can, in fact, achieve the optimal tradeoffs. In particular, separate encoding can achieve the MSR point $(7/18,11/36)$ and the MBR point $(8/15,8/45)$. In the same figure, the outer bounds \eqref{eq:B1} and \eqref{eq:B2} have also been plotted. As illustrated, they intersect precisely at the MBR point $(8/15,8/45)$. Notice that for this example at least, the outer bound \eqref{eq:B2} is tight {\em only} at the MBR point.

\begin{figure*}[t!]
\centering
\includegraphics[width=0.58\linewidth,draft=false]{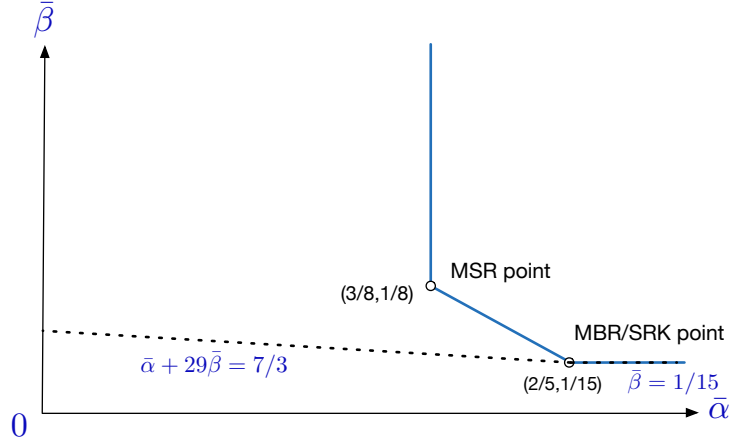}
\caption{The optimal tradeoff curve between the normalized storage-capacity $\bar{\alpha}$ and repair-bandwidth $\bar{\beta}$ for the $(7,6,6,1)$ SRC problem \cite{Shao-IT17}. The outer bounds \eqref{eq:B5} and \eqref{eq:B6} are evaluated as $\bar{\beta}\geq 1/15$ and $\bar{\alpha}+29\bar{\beta} \geq 7/3$, respectively. When set as equalities, they intersect precisely at the MBR/SRK point $(2/5,1/15)$.}
\label{fig2}
\end{figure*}

\section{Main Results}\label{sec:Main}
Our main result of the paper is to show that the optimality of separate coding in terms of achieving the MBR point of the normalized storage-capacity repair-bandwidth tradeoff region extends more generally from the MDC-R problem to the MDC-SR problem. The results are summarized in the following theorem.

\begin{thm}\label{thm}
For the general MDC-SR problem, any achievable normalized message-rate storage-capacity repair-bandwidth tuple $(\bar{B}_{\ell+1},\ldots,\bar{B}_d,\bar{\alpha},\bar{\beta}) \in \mathcal{R}_{n,d,\ell}$ must satisfy:
\begin{align}
\bar{\beta} & \geq \sum_{j=\ell+1}^{d}T_{d,j,\ell}^{-1}\bar{B}_j \label{eq:B3}\\
\mbox{and} \quad \bar{\alpha}+(d(d-\ell)-\ell)\bar{\beta} & \geq (d-\ell)(d+1)\sum_{j=\ell+1}^{d}T_{d,j,\ell}^{-1}\bar{B}_j\label{eq:B4}
\end{align}
where $T_{d,k,\ell}:=\sum_{t=\ell+1}^{k}(d+1-t)$. When set as equalities, the intersection of \eqref{eq:B3} and \eqref{eq:B4} is given by:
\begin{align*}
\left(\bar{\alpha},\bar{\beta}\right)
&=\left(d\sum_{j=\ell+1}^{d}T_{d,j,\ell}^{-1}\bar{B}_j,\sum_{j=\ell+1}^{d}T_{d,j,\ell}^{-1}\bar{B}_j\right).
\end{align*}
For any $j\in [\ell+1:d]$, the SRK point for the $(n,j,d,\ell)$ SRC problem can be written as \cite{Shah-Globecom11}: 
\begin{align}
(dT_{d,j,\ell}^{-1},T_{d,j,\ell}^{-1}) \in \mathcal{R}_{n,j,d,\ell}.\label{eq:SRK}
\end{align}
We may thus conclude immediately from \eqref{eq:SC} that separate coding can achieve the MBR point for the general MDC-SR problem.
\end{thm}

The following corollary follows immediately from Theorem~\ref{thm} by setting $\bar{B}_j=0$ for all $j \neq k$.

\begin{coro}
For the general SRC problem, any achievable normalized storage-capacity repair-bandwidth tuple $(\bar{\alpha},\bar{\beta}) \in \mathcal{R}_{n,k,d,\ell}$ must satisfy:
\begin{align}
\bar{\beta} & \geq T_{d,k,\ell}^{-1} \label{eq:B5}\\
\mbox{and} \quad \bar{\alpha}+(d(d-\ell)-\ell)\bar{\beta} & \geq (d-\ell)(d+1)T_{d,k,\ell}^{-1}.\label{eq:B6}
\end{align}
When set as equalities, the intersection of \eqref{eq:B5} and \eqref{eq:B6} is precisely the SRK point \eqref{eq:SRK} (with $j=k$), showing that the SRK point is, in fact, the MBR point of the achievable normalized storage-capacity repair-bandwidth tradeoff region for the general SRC problem.
\end{coro}

While the outer bound \eqref{eq:B5} is known \cite{Pawar-ISIT10,Pawar-IT11,Shao-IT17}, the outer bound \eqref{eq:B6} is new to the best of our knowledge. Fig.~\ref{fig2} shows the optimal tradeoff curve between the normalized storage-capacity and repair-bandwidth for the $(7,6,6,1)$ SRC problem. Notice that for this example, the SRK point $(2/5,1/15)$ is, in fact, the MBR point even though the tradeoff region has two corner points. In the same figure, the outer bunds \eqref{eq:B5} and \eqref{eq:B6} have also been plotted. As illustrated, when set as equalities, they intersect precisely at the MBR/SRK point $(2/5,1/15)$. Notice that for this example at least, the outer bound \eqref{eq:B6} is tight {\em only} at the MBR/SRK point. 

As a final remark, we mention here that when $\ell=0$, the outer bound \eqref{eq:B3} is reduced to \eqref{eq:B1} for the $(n,d)$ MDC-R problem by the fact that $T_{n,d,0}=T_{n,d}$. However, when $\ell=0$, the outer bound \eqref{eq:B4} is reduced to:
\begin{align}
\bar{\alpha}+d^2\bar{\beta} & \geq d(d+1)\sum_{j=1}^{d}T_{d,j}^{-1}\bar{B}_j\label{eq:B7}
\end{align}
which is {\em weaker} than the outer bound \eqref{eq:B2} by the fact that $d^2 >\frac{d(d-1)}{2}$. Fig.~\ref{fig1} shows the outer bound \eqref{eq:B7} for the $(4,3)$ MDC-R problem with $(\bar{B}_1,\bar{B}_2,\bar{B}_3)=(0,1/3,2/3)$. As illustrated, \eqref{eq:B7} is weaker than \eqref{eq:B2}, and both are {\em only} tight at the MBR point $(8/15,8/45)$.

\section{Proof of the Main Results}\label{sec:Proof}
Let us first outline the main ingredients for proving the outer bounds \eqref{eq:B3} and \eqref{eq:B4}.

\begin{itemize}
\item[1)] {\em Total number of nodes.} To prove the outer bounds \eqref{eq:B3} and \eqref{eq:B4}, let us first note that these bounds are {\em independent} of the total number of storage nodes $n$ in the system. Therefore, in our proof, we only need to consider the cases where $n=d+1$. For the cases where $n>d+1$, since any subsystem consisting of $d+1$ out of the total $n$ storage nodes must give rise to a $(d+1,d,\ell)$ MDC-SR problem. Therefore, these {\em outer} bounds must apply as well. When $n=d+1$, any repair group $\mathcal{B}$ of size $d$ is uniquely determined by the node $j$ to be repaired, i.e., $\mathcal{B}=[1:n]\setminus\{j\}$, and hence can be dropped from the notation $\mathsf{S}^{\mathcal{B}}_{i \rightarrow j}$ without causing any confusion. 
\item[2)] {\em Code symmetry.} Due to the built-in {\em symmetry} of the problem, to prove the outer bounds \eqref{eq:B3}, and \eqref{eq:B4}, we only need to consider the so-called {\em symmetrical} codes \cite{Tian-JSAC13,Zhang-TCOM2017} for which the joint entropy of any subset of random variables from
\begin{align*}
&\left((\mathsf{M}_1,\ldots,\mathsf{M}_d),\mathsf{K},\right.\\
&\left.\hspace{30pt}(\mathsf{W}_i:i\in[1:n]),(\mathsf{S}_{i \rightarrow j}: i,j\in[1:n],i\neq j)\right)
\end{align*} 
remains {\em unchanged} under {\em any} permutation over the {\em storage-node} indices. 
\item[3)] {\em Key collections of random variables.} Focusing on the symmetrical $(n=d+1,d,N_1,\ldots,N_d,K,T,S)$ codes, the following collections of random variables play a key role in our proof:
\begin{align*}
& \mathsf{M}_{\mathcal{A}} := (\mathsf{M}_i: i \in \mathcal{A}), \quad \mathcal{A}\subseteq [1:d]\\
& \mathsf{M}^{(m)} := \mathsf{M}_{[1:m]}, \quad m\in[1:d]\\
&\mathsf{W}_{\mathcal{A}} :=\left(\mathsf{W}_i:i\in \mathcal{A}\right), \quad \mathcal{A}\subseteq [1:n]\\
&\mathsf{S}_{i\rightarrow\mathcal{B}} := \left(\mathsf{S}_{i \rightarrow j}: j\in\mathcal{B}\right),\quad i\in [1:n], \; \mathcal{B}\subseteq [1:n]\setminus \{i\}\\
&\mathsf{S}_{\mathcal{B}\rightarrow j} := \left(\mathsf{S}_{i \rightarrow j}: \quad i\in\mathcal{B}\right), j\in[1:n],\; \mathcal{B}\subseteq [1:n]\setminus \{j\}\\
&\mathsf{S}_{\rightarrow j} := \mathsf{S}_{[1:j-1]\cup[j+1:n]\rightarrow j}, \quad j\in[1:n]\\
&\mathsf{S}_{\rightarrow \mathcal{B}} :=\left(\mathsf{S}_{\rightarrow j}:j\in\mathcal{B}\right), \quad \mathcal{B}\subseteq [1:n]\\
&\underline{\mathsf{S}}_{\rightarrow j} := \mathsf{S}_{[1:j-1]\rightarrow j}, \quad j\in[1:n]\\
&\underline{\mathsf{S}}_{\rightarrow \mathcal{B}} :=(\underline{\mathsf{S}}_{\rightarrow j}:j\in\mathcal{B}), \quad \mathcal{B}\subseteq [1:n]\\
&\overline{\mathsf{S}}_{\rightarrow j} := \mathsf{S}_{[j+1:n]\rightarrow j}, \quad j\in[1:n]\\
&\overline{\mathsf{S}}_{\rightarrow \mathcal{B}} :=(\overline{\mathsf{S}}_{\rightarrow j}:j\in\mathcal{B}), \quad \mathcal{B}\subseteq [1:n]\\
&\mathsf{U}^{(t,s)} :=(\mathsf{W}_{[1:t]},\overline{\mathsf{S}}_{\rightarrow[t+1:s]}), \quad s\in[1:n], \; t\in[0:s]\\
&\mathsf{U}^{(s)} := \mathsf{U}^{(0,s)}.
\end{align*}
These collections of random variables have also been used in \cite{Shao-IT17,Shao-CISS16}.
\end{itemize}

Note that if we consider representing the collection of the random variables $\{\mathsf{S}_{i \rightarrow j}\}$ as an $n$-by-$n$ matrix and write $\{\mathsf{W}_i\}$ on the diagonal of this matrix, then $\mathsf{U}^{(t,s)}$ is the collection of these random variables with an {\em upper triangular} pattern. An important part of the proof is to understand the relations between different $\mathsf{U}^{(t,s)}$'s (conditioned on a subset of messages) and then use them to derive the desired converse results. We shall discuss this next.

\subsection{Technical Lemmas}

\begin{lemma}\label{lemma1}
For any $(n=d+1,d,N_1,\ldots,N_d,K,T,S)$ code that satisfies the node regeneration requirement \eqref{eq:NodeRegen}, $(\underline{\mathsf{S}}_{\rightarrow[t+1:s]},\mathsf{W}_{[t+1:s]})$ is a function of $\mathsf{U}^{(t,s)}$ for any $s \in [1:n]$ and $t\in[0:s-1]$. 
\end{lemma}

\begin{IEEEproof}
Fix $s \in [1:n]$ and $t\in[0:s-1]$. Let us first note that $\underline{\mathsf{S}}_{\rightarrow t+1}$ is a function of $\mathsf{W}_{[1:t]}$. As a result, $\mathsf{S}_{\rightarrow t+1}=(\underline{\mathsf{S}}_{\rightarrow t+1},\overline{\mathsf{S}}_{\rightarrow t+1})$ is a function of $\mathsf{U}^{(t,s)}$. It thus follows immediately from the node regeneration requirement \eqref{eq:NodeRegen} that $\mathsf{W}_{t+1}$ is a function of $\mathsf{U}^{(t,s)}$. Similarly and inductively, it can be shown that $(\underline{\mathsf{S}}_{\rightarrow j},\mathsf{W}_j)$ is a function of $\mathsf{U}^{(t,s)}$ for all $j\in [t+2:s]$. This completes the proof of the lemma.
\end{IEEEproof}

The above lemma demonstrates the ``compactness" of $\mathsf{U}^{(t,s)}$ and has a number of direct consequences. For example, for any fixed $s \in [1:n]$, it is clear from Lemma~\ref{lemma1} that $\mathsf{U}^{(t_2,s)}$ is a function of $\mathsf{U}^{(t_1,s)}$ and hence $H(\mathsf{U}^{(t_2,s)}) \leq H(\mathsf{U}^{(t_1,s)})$ for any $0 \leq t_1 \leq t_2 \leq s-1$. 

The following lemma describes an ``exchange" relation between $\mathsf{U}^{(i,m)}$ and $\mathsf{U}^{(i',j)}$, which plays the key role in proving the outer bounds \eqref{eq:B1} and \eqref{eq:B2}. The proof is rather long and is deferred to the Appendix to enhance the flow of the paper.

\begin{lemma}[Exchange lemma]
\label{lemma:exchange}
For any symmetrical $(n=d+1,d,N_1,\ldots,N_d,K,T,S)$ code that satisfies the node regeneration requirement \eqref{eq:NodeRegen}, we have
\begin{align}
& \frac{d+1-j}{d-m}H(\mathsf{U}^{(i,m)}|\mathsf{M}^{(m)})+H(\mathsf{U}^{(i',j)}|\mathsf{M}^{(m)})\nonumber\\
& \hspace{15pt} \geq \frac{d+1-j}{d-m}H(\mathsf{U}^{(i,m+1)}|\mathsf{M}^{(m)})+H(\mathsf{U}^{(i',j-1)}|\mathsf{M}^{(m)})\label{eq:exchange}
\end{align}
for any $m\in[1:d-1]$, $i\in[0:m-1]$, $i'\in[0:i]$, and $j\in[i'+1:m-i+i'+1]$.
\end{lemma}

\begin{coro}\label{coro1}
For any symmetrical $(n=d+1,d,N_1,\ldots,N_d,K,T,S)$ code that satisfies the node regeneration requirement \eqref{eq:NodeRegen}, we have
\begin{align}
&T_{d,m,\ell}^{-1}H(\mathsf{U}^{(m)}|\mathsf{M}^{(m)})\geq T_{d,m+1,\ell}^{-1}H(\mathsf{U}^{(m+1)}|\mathsf{M}^{(m)})+\nonumber\\
& \hspace{110pt} (T_{d,m,\ell}^{-1}-T_{d,m+1,\ell}^{-1})H(\mathsf{U}^{(\ell)}|\mathsf{M}^{(m)})\label{eq:coro1}
\end{align}
for any $\ell\in[0:d-1]$ and $m\in[\ell+1:d-1]$.
\end{coro}

\begin{IEEEproof}
Fix $\ell\in[0:d-1]$ and $m\in[\ell+1:d-1]$. Setting $i=i'=0$ in \eqref{eq:exchange}, we have
\begin{align}
& \frac{d+1-j}{d-m}H(\mathsf{U}^{(m)}|\mathsf{M}^{(m)})+H(\mathsf{U}^{(j)}|\mathsf{M}^{(m)})\nonumber\\
& \hspace{10pt} \geq \frac{d+1-j}{d-m}H(\mathsf{U}^{(m+1)}|\mathsf{M}^{(m)})+H(\mathsf{U}^{(j-1)}|\mathsf{M}^{(m)})\label{eq:NewExchange1}
\end{align}
for any $j\in[1:m+1]$. Add the inequalities \eqref{eq:NewExchange1} for $j\in[\ell+1:m]$ and cancel the common term $\sum_{j=\ell+1}^{m-1}H(\mathsf{U}^{(j)}|\mathsf{M}^{(m)})$ from both sides. We have
\begin{align*}
&\frac{T_{d,m,\ell}}{d-m}H(\mathsf{U}^{(m)}|\mathsf{M}^{(m)})+H(\mathsf{U}^{(m)}|\mathsf{M}^{(m)})\nonumber\\
& \hspace{10pt}\geq \frac{T_{d,m,\ell}}{d-m}H(\mathsf{U}^{(m+1)}|\mathsf{M}^{(m)})+H(\mathsf{U}^{(\ell)}|\mathsf{M}^{(m)})
\end{align*}
which can be equivalently written as 
\begin{align}
&\frac{T_{d,m+1,\ell}}{d-m}H(\mathsf{U}^{(m)}|\mathsf{M}^{(m)})\nonumber\\
& \hspace{30pt} \geq \frac{T_{d,m,\ell}}{d-m}H(\mathsf{U}^{(m+1)}|\mathsf{M}^{(m)})+ H(\mathsf{U}^{(\ell)}|\mathsf{M}^{(m)})\label{eq:Temp4}
\end{align}
by the fact that $T_{d,m,\ell}+(d-m)=T_{d,m+1,\ell}$. Multiplying both sides of \eqref{eq:Temp4} by 
$$\frac{d-m}{T_{d,m+1,\ell}T_{d,m,\ell}}=T_{d,m,\ell}^{-1}-T_{d,m+1,\ell}^{-1}$$ 
completes the proof of \eqref{eq:coro1}.
\end{IEEEproof}

\begin{coro}\label{coro2}
For any symmetrical $(n=d+1,d,N_1,\ldots,N_d,K,T,S)$ code that satisfies the node regeneration requirement \eqref{eq:NodeRegen}, we have
\begin{align}
H&(\mathsf{U}^{(1,m)}|\mathsf{M}^{(m)})+(d-m)T_{d,m,\ell}^{-1}H(\mathsf{U}^{(m)}|\mathsf{M}^{(m)})\nonumber\\
&\geq H(\mathsf{U}^{(1,m+1)}|\mathsf{M}^{(m)})+(d-m)T_{d,m,\ell}^{-1}H(\mathsf{U}^{(\ell)}|\mathsf{M}^{(m)})\label{eq:coro2}
\end{align}
for any $\ell\in[0:d-1]$ and $m\in[\ell+1:d-1]$.
\end{coro}

\begin{IEEEproof}
Fix $\ell\in[0:d-1]$ and $m\in[\ell+1:d-1]$. Set $i=1$ and $i'=0$ in \eqref{eq:exchange}. We have
\begin{align}
& \frac{d+1-j}{d-m}H(\mathsf{U}^{(1,m)}|\mathsf{M}^{(m)})+H(\mathsf{U}^{(j)}|\mathsf{M}^{(m)})\nonumber\\
& \hspace{10pt}\geq \frac{d+1-j}{d-m}H(\mathsf{U}^{(1,m+1)}|\mathsf{M}^{(m)})+H(\mathsf{U}^{(j-1)}|\mathsf{M}^{(m)})\label{eq:NewExchange2}
\end{align}
for any $j\in[1:m]$. Add the inequalities \eqref{eq:NewExchange2} for $j\in[\ell+1:m]$ and cancel the common term $\sum_{j=\ell+1}^{m-1}H(\mathsf{U}^{(j)}|\mathsf{M}^{(m)})$ from both sides. We have
\begin{align}
&\frac{T_{d,m,\ell}}{d-m}H(\mathsf{U}^{(1,m)}|\mathsf{M}^{(m)})+H(\mathsf{U}^{(m)}|\mathsf{M}^{(m)})\nonumber\\
& \hspace{10pt}\geq \frac{T_{d,m,\ell}}{d-m}H(\mathsf{U}^{(1,m+1)}|\mathsf{M}^{(m)})+H(\mathsf{U}^{(\ell)}|\mathsf{M}^{(m)}).\label{eq:NewExchange3}
\end{align}
Multiplying both sides of \eqref{eq:NewExchange3} by $(d-m)T_{d,m,\ell}^{-1}$ completes the proof of \eqref{eq:coro2}.
\end{IEEEproof}

\subsection{The Proof}
Consider a symmetrical $(n=d+1,d,1,\ldots,1,N_{\ell+1},\ldots,N_d,K,T,S)$ regenerating code that satisfies the rate normalization requirement \eqref{eq:Rate}, the message recovery requirement \eqref{eq:MessageRecovery}, the node regeneration requirement \eqref{eq:NodeRegen}, and the repair secrecy requirement \eqref{eq:RepairSecrecy}. Let us first prove a few intermediate results. The outer bounds \eqref{eq:B3} and \eqref{eq:B4} will then follow immediately.

\begin{prop}\label{prop1}
\begin{align}
&\frac{1}{d-\ell}H(\mathsf{U}^{(\ell+1)}) \geq \sum_{j=\ell+1}^{m}T_{d,j,\ell}^{-1}B_j+\nonumber\\
& \hspace{20pt} T_{d,m,\ell}^{-1}H(\mathsf{U}^{(m)}|\mathsf{M}_{[\ell+1:m]})+\left(\frac{1}{d-\ell}-T_{d,m,\ell}^{-1}\right)H(\mathsf{U}^{(\ell)})\label{eq:EG}
\end{align}
for any $m\in [\ell+1:d]$. Consequently,
\begin{align}
\frac{1}{d-\ell}H(\mathsf{U}^{(\ell+1)}) \geq \sum_{j=\ell+1}^{d}T_{d,j,\ell}^{-1}B_j+\frac{1}{d-\ell}H(\mathsf{U}^{(\ell)}).
\label{eq:prop1}
\end{align}
\end{prop}

\begin{IEEEproof}
To see \eqref{eq:EG}, consider proof by induction. For the base case with $m=\ell+1$, we have
\begin{align*}
\frac{1}{d-\ell}H(\mathsf{U}^{(\ell+1)})
& \stackrel{(a)}{=} \frac{1}{d-\ell}H(\mathsf{U}^{(\ell+1)},\mathsf{M}_{\ell+1})\\
& \stackrel{(b)}{=} \frac{1}{d-\ell}\left(H(\mathsf{M}_{\ell+1})+H(\mathsf{U}^{(\ell+1)}|\mathsf{M}_{\ell+1})\right)\\ 
& \stackrel{(c)}{=} \frac{1}{d-\ell}\left(B_{\ell+1}+H(\mathsf{U}^{(\ell+1)}|\mathsf{M}_{\ell+1})\right)\\ 
& \stackrel{(d)}{=} T_{d,\ell+1,\ell}^{-1}B_{\ell+1}+T_{d,\ell+1,\ell}^{-1}H(\mathsf{U}^{(\ell+1)}|\mathsf{M}_{\ell+1})
\end{align*}
where $(a)$ follows from the fact that $\mathsf{M}_{\ell+1}$ is a function of $\mathsf{W}_{[1:\ell+1]}$, which is a function of $\mathsf{U}^{(\ell+1)}$ by Lemma~\ref{lemma1}; $(b)$ follows from the chain rule for entropy; $(c)$ follows from the fact that $H(\mathsf{M}_{\ell+1})=B_{\ell+1}$; and $(d)$ follows from the fact that $T_{d,\ell+1,\ell}=d-\ell$. Assuming that \eqref{eq:EG} holds for some $m\in [\ell+1:d-1]$, we have
\begin{align*}
& \frac{1}{d-\ell}H(\mathsf{U}^{(\ell+1)})\\
& \stackrel{(a)}{\geq} \sum_{j=\ell+1}^{m}T_{d,j,\ell}^{-1}B_j+T_{d,m,\ell}^{-1}H(\mathsf{U}^{(m)}|\mathsf{M}_{[\ell+1:m]})+\\
& \hspace{20pt} \left(\frac{1}{d-\ell}-T_{d,m,\ell}^{-1}\right)H(\mathsf{U}^{(\ell)})\\
& \stackrel{(b)}{\geq} \sum_{j=\ell+1}^{m}T_{d,j,\ell}^{-1}B_j+T_{d,m+1,\ell}^{-1}H(\mathsf{U}^{(m+1)}|\mathsf{M}_{[\ell+1:m]})+\\
& \hspace{20pt} \left(\frac{1}{d-\ell}-T_{d,m+1,\ell}^{-1}\right)H(\mathsf{U}^{(\ell)})\\
& \stackrel{(c)}{\geq} \sum_{j=\ell+1}^{m}T_{d,j,\ell}^{-1}B_j+T_{d,m+1,\ell}^{-1}H(\mathsf{U}^{(m+1)},\mathsf{M}_{m+1}|\mathsf{M}_{[\ell+1:m]})+\\
& \hspace{20pt} \left(\frac{1}{d-\ell}-T_{d,m+1,\ell}^{-1}\right)H(\mathsf{U}^{(\ell)})\\
& \stackrel{(d)}{=} \sum_{j=\ell+1}^{m}T_{d,j,\ell}^{-1}B_j+T_{d,m+1,\ell}^{-1}H(\mathsf{M}_{m+1}|\mathsf{M}_{[\ell+1:m]})+\\
& \hspace{20pt} T_{d,m+1,\ell}^{-1}H(\mathsf{U}^{(m+1)}|\mathsf{M}_{[\ell+1:m+1]})+\\
& \hspace{20pt} \left(\frac{1}{d-\ell}-T_{d,m+1,\ell}^{-1}\right)H(\mathsf{U}^{(\ell)})\\
& \stackrel{(e)}{=} \sum_{j=\ell+1}^{m}T_{d,j,\ell}^{-1}B_j+T_{d,m+1,\ell}^{-1}B_{m+1}+\\
& \hspace{20pt} T_{d,m+1,\ell}^{-1}H(\mathsf{U}^{(m+1)}|\mathsf{M}_{[\ell+1:m+1]})+\\
& \hspace{20pt} \left(\frac{1}{d-\ell}-T_{d,m+1,\ell}^{-1}\right)H(\mathsf{U}^{(\ell)})\\
& = \sum_{j=\ell+1}^{m+1}T_{d,j,\ell}^{-1}+T_{d,m+1,\ell}^{-1}H(\mathsf{U}^{(m+1)}|\mathsf{M}_{[\ell+1:m+1]})+\\
& \hspace{20pt} \left(\frac{1}{d-\ell}-T_{d,m+1,\ell}^{-1}\right)H(\mathsf{U}^{(\ell)})
\end{align*} 
where $(a)$ follows from the induction assumption; $(b)$ follows from Corollary~\ref{coro1}; $(c)$ follows from the fact that $\mathsf{M}_{m+1}$ is a function of $\mathsf{W}_{[1:m+1]}$, which is is a function of $\mathsf{U}^{(m+1)}$ by Lemma~\ref{lemma1}; $(d)$ follows from the chain rule for entropy; and $(e)$ follows from the facts that $\mathsf{M}_{m+1}$ is independent of $\mathsf{M}_{[\ell+1:m]}$ and that $H(\mathsf{M}_{m+1})=B_{m+1}$. This completes the induction step and hence the proof of \eqref{eq:EG}.

To see \eqref{eq:prop1}, simply set $m=d$ in \eqref{eq:EG}. We have
\begin{align}
&\frac{1}{d-\ell}H(\mathsf{U}^{(\ell+1)}) \geq \sum_{j=\ell+1}^{d}T_{d,j,\ell}^{-1}B_j+\nonumber\\
& \hspace{40pt} T_{d,d,\ell}^{-1}H(\mathsf{U}^{(d)}|\mathsf{M}_{[\ell+1:d]})+\left(\frac{1}{d-\ell}-T_{d,d,\ell}^{-1}\right)H(\mathsf{U}^{(\ell)}).\label{eq:EG2}
\end{align}
Note that
\begin{align}
H(\mathsf{U}^{(d)}|\mathsf{M}_{[\ell+1:d]}) \geq H(\mathsf{U}^{(\ell)}|\mathsf{M}_{[\ell+1:d]}) = H(\mathsf{U}^{(\ell)})\label{eq:EG3}
\end{align}
where the last equality follows from the fact that $I(\mathsf{U}^{(\ell)};\mathsf{M}_{[\ell+1:d]})=0$ by the repair secrecy requirement \eqref{eq:RepairSecrecy}. Substituting \eqref{eq:EG3} into \eqref{eq:EG2} completes the proof of \eqref{eq:prop1}.
\end{IEEEproof}

\begin{prop}\label{prop2}
\begin{align}
H(\mathsf{S}_{d+1\rightarrow[1:\ell]})+(d(d-\ell)-\ell)\beta+dH(\mathsf{U}^{(\ell)})\geq
dH(\mathsf{U}^{(\ell+1)}).\label{eq:prop2}
\end{align}
\end{prop}

\begin{IEEEproof}
First note that for any $m \in [1:\ell]$, we have
\begin{align}
H&(\mathsf{S}_{d+1\rightarrow[1:m]})+H(\mathsf{U}^{(\ell)})\nonumber\\
& \stackrel{(a)}{=} H(\mathsf{S}_{d+1\rightarrow[1:m-1]\cup\{\ell+1\}})+H(\mathsf{U}^{(\ell)})\nonumber\\
& \stackrel{(b)}{\geq} H(\mathsf{S}_{d+1\rightarrow[1:m-1]})+H(\mathsf{U}^{(\ell)},\mathsf{S}_{d+1\rightarrow\ell+1})\label{eq:STE}
\end{align}
where $(a)$ follows from the fact that $H(\mathsf{S}_{d+1\rightarrow[1:m]})=H(\mathsf{S}_{d+1\rightarrow[1:m-1]\cup\{\ell+1\}})$ due to the symmetrical code that we consider, and $(b)$ follows from the submodularity of the entropy function. Add \eqref{eq:STE} over $m\in[1:\ell]$ and cancel $\sum_{m=1}^{\ell-1}H(\mathsf{S}_{d+1\rightarrow[1:m]})$ from both sides. We have
\begin{align}
H(\mathsf{S}_{d+1\rightarrow[1:\ell]})+\ell H(\mathsf{U}^{(\ell)}) \geq \ell H(\mathsf{U}^{(\ell)},\mathsf{S}_{d+1\rightarrow\ell+1}).\label{eq:STE2}
\end{align}
It follows that
\begin{align*}
H&(\mathsf{S}_{d+1\rightarrow[1:\ell]})+(d(d-\ell)-\ell)\beta+dH(\mathsf{U}^{(\ell)})\\
& = \left(H(\mathsf{S}_{d+1\rightarrow[1:\ell]})+\ell H(\mathsf{U}^{(\ell)})\right)+\\
& \hspace{20pt} (d(d-\ell)-\ell)\beta+(d-\ell)H(\mathsf{U}^{(\ell)})\\
& \stackrel{(a)}{\geq} \ell H(\mathsf{U}^{(\ell)},\mathsf{S}_{d+1\rightarrow\ell+1})+(d(d-\ell)-\ell)\beta+(d-\ell)H(\mathsf{U}^{(\ell)})\\
& = \ell\left((d-\ell-1)\beta+H(\mathsf{U}^{(\ell)},\mathsf{S}_{d+1\rightarrow\ell+1})\right)+\\
& \hspace{20pt} (d-\ell)\left((d-\ell)\beta+H(\mathsf{U}^{(\ell)})\right)\\
& \stackrel{(b)}{\geq} \ell\left(H(\mathsf{S}_{[\ell+2:d]\rightarrow \ell+1})+H(\mathsf{U}^{(\ell)},\mathsf{S}_{d+1\rightarrow\ell+1})\right)+\\
& \hspace{20pt} (d-\ell)\left(H(\overline{\mathsf{S}}_{\rightarrow \ell+1})+H(\mathsf{U}^{(\ell)})\right)\\
& \stackrel{(c)}{\geq} \ell H(\mathsf{U}^{(\ell+1)})+(d-\ell)H(\mathsf{U}^{(\ell+1)})\\
& = dH(\mathsf{U}^{(\ell+1)})
\end{align*}
where $(a)$ follows from \eqref{eq:STE2}; $(b)$ follows from the facts that $H(\mathsf{S}_{[\ell+2:d]\rightarrow \ell+1}) \leq (d-\ell-1)\beta$ and that $H(\overline{\mathsf{S}}_{\rightarrow \ell+1}) \leq (d-\ell)\beta$; and $(c)$ follows from the facts that $H(\mathsf{S}_{[\ell+2:d]\rightarrow \ell+1})+H(\mathsf{U}^{(\ell)},\mathsf{S}_{d+1\rightarrow\ell+1}) \geq H(\mathsf{U}^{(\ell+1)})$ and that $H(\overline{\mathsf{S}}_{\rightarrow \ell+1})+H(\mathsf{U}^{(\ell)})\geq H(\mathsf{U}^{(\ell+1)})$ by the union bound on entropy. This completes the proof of the proposition.
\end{IEEEproof}

\begin{prop}\label{prop3}
\begin{align}
&H(\mathsf{U}^{(1,m)})+\frac{d-m}{d-\ell}H(\mathsf{U}^{(\ell+1)})\nonumber\\
& \geq (d-m)\sum_{j=\ell+1}^{m}T_{d,j,\ell}^{-1}B_j+H(\mathsf{U}^{(1,m+1)})+\frac{d-m}{d-\ell}H(\mathsf{U}^{(\ell)})\label{eq:JH}
\end{align}
for any $m\in[\ell+1,d-1]$. Consequently,
\begin{align}
& H(\mathsf{U}^{(1,\ell+1)})+\frac{T_{d,d,\ell+1}}{d-\ell}H(\mathsf{U}^{(\ell+1)})\nonumber\\
& \hspace{20pt} \geq T_{d,d,\ell}\sum_{j=\ell+1}^{d}T_{d,j,\ell}^{-1}B_j+\frac{T_{d,d,\ell}}{d-\ell}H(\mathsf{U}^{(\ell)}).\label{eq:prop3}
\end{align}
\end{prop}

\begin{IEEEproof}
To see \eqref{eq:JH}, note that for any $m\in[\ell+1,d-1]$, we have
\begin{align*}
&H(\mathsf{U}^{(1,m)}|\mathsf{M}_{[\ell+1:m]})+\frac{d-m}{d-\ell}H(\mathsf{U}^{(\ell+1)})\\
& \stackrel{(a)}{\geq} H(\mathsf{U}^{(1,m)}|\mathsf{M}_{[\ell+1:m]})+(d-m)\left(\sum_{j=\ell+1}^{m}T_{d,j,\ell}^{-1}B_j+\right.\\
& \hspace{20pt} \left.T_{d,m,\ell}^{-1}H(\mathsf{U}^{(m)}|\mathsf{M}_{[\ell+1:m]})+\left(\frac{1}{d-\ell}-T_{d,m,\ell}^{-1}\right)H(\mathsf{U}^{(\ell)})\right)\\
& = H(\mathsf{U}^{(1,m)}|\mathsf{M}_{[\ell+1:m]})+(d-m)T_{d,m,\ell}^{-1}H(\mathsf{U}^{(m)}|\mathsf{M}_{[\ell+1:m]})+\\
& \hspace{20pt} (d-m)\left(\sum_{j=\ell+1}^{m}T_{d,j,\ell}^{-1}B_j+\left(\frac{1}{d-\ell}-T_{d,m,\ell}^{-1}\right)H(\mathsf{U}^{(\ell)})\right)\\
& \stackrel{(b)}{\geq} H(\mathsf{U}^{(1,m+1)}|\mathsf{M}_{[\ell+1:m]})+(d-m)T_{d,m,\ell}^{-1}H(\mathsf{U}^{(\ell)}|\mathsf{M}_{[\ell+1:m]})+\\
& \hspace{20pt} (d-m)\left(\sum_{j=\ell+1}^{m}T_{d,j,\ell}^{-1}B_j+\left(\frac{1}{d-\ell}-T_{d,m,\ell}^{-1}\right)H(\mathsf{U}^{(\ell)})\right)\\
& \stackrel{(c)}{=} H(\mathsf{U}^{(1,m+1)}|\mathsf{M}_{[\ell+1:m]})+(d-m)T_{d,m,\ell}^{-1}H(\mathsf{U}^{(\ell)})+\\
& \hspace{20pt} (d-m)\left(\sum_{j=\ell+1}^{m}T_{d,j,\ell}^{-1}B_j+\left(\frac{1}{d-\ell}-T_{d,m,\ell}^{-1}\right)H(\mathsf{U}^{(\ell)})\right)\\
& = H(\mathsf{U}^{(1,m+1)}|\mathsf{M}_{[\ell+1:m]})+(d-m)\sum_{j=\ell+1}^{m}T_{d,j,\ell}^{-1}B_j+\\
& \hspace{20pt} \frac{d-m}{d-\ell}H(\mathsf{U}^{(\ell)})
\end{align*}
where $(a)$ follows from \eqref{eq:EG} of Proposition~\ref{prop1}; $(b)$ follows from Corollary~\ref{coro2}; and $(c)$ follows from the fact that $I(\mathsf{U}^{(\ell)};\mathsf{M}_{[\ell+1:m]})=0$ due to the repair secrecy requirement \eqref{eq:RepairSecrecy}. Adding $H(\mathsf{M}_{[\ell+1:m]})$ to both sides and using the facts that
\begin{align*}
&H(\mathsf{U}^{(1,m)}|\mathsf{M}_{[\ell+1:m]})+H(\mathsf{M}_{[\ell+1:m]})\\
&\hspace{40pt}=H(\mathsf{U}^{(1,m)},\mathsf{M}_{[\ell+1:m]})\stackrel{(a)}{=}H(\mathsf{U}^{(1,m)})
\end{align*}
and that
\begin{align*}
&H(\mathsf{U}^{(1,m+1)}|\mathsf{M}_{[\ell+1:m]})+H(\mathsf{M}_{[\ell+1:m]})\\
&\hspace{40pt}=H(\mathsf{U}^{(1,m+1)},\mathsf{M}_{[\ell+1:m]})\stackrel{(b)}{=}H(\mathsf{U}^{(1,m+1)})
\end{align*}
complete the proof of \eqref{eq:JH}. Here, $(a)$ and $(b)$ are due to the facts that $\mathsf{M}_{[\ell+1:m]}$ is a function of $\mathsf{W}_{[1:m]}$, which is a function of both $\mathsf{U}^{(1,m)}$ and $\mathsf{U}^{(1,m+1)}$ by Lemma~\ref{lemma1}.

To see \eqref{eq:prop3}, add \eqref{eq:JH} over $m\in[\ell+1:d-1]$ and cancel $\sum_{m=\ell+2}^{d-1}H(\mathsf{U}^{(1,m)})$ from both sides of the inequality. We have
\begin{align}
& H(\mathsf{U}^{(1,\ell+1)})+\frac{T_{d,d,\ell+1}}{d-\ell}H(\mathsf{U}^{(\ell+1)})\nonumber\\
& \hspace{40pt} \geq \sum_{m=\ell+1}^{d-1}\left((d-m)\sum_{j=\ell+1}^{m}T_{d,j,\ell}^{-1}B_j\right)+\nonumber\\
& \hspace{80pt} H(\mathsf{U}^{(1,d)})+\frac{T_{d,d,\ell+1}}{d-\ell}H(\mathsf{U}^{(\ell)}).\label{eq:JH1}
\end{align}
Note that
\begin{align}
& \sum_{m=\ell+1}^{d-1}\left((d-m)\sum_{j=\ell+1}^{m}T_{d,j,\ell}^{-1}B_j\right)\nonumber\\
& \hspace{10pt} = \sum_{j=\ell+1}^{d-1}T_{d,j,\ell}^{-1}B_j\left(\sum_{m=j}^{d-1}(d-m)\right)
= \sum_{j=\ell+1}^{d-1}T_{d,j,\ell}^{-1}T_{d,d,j}B_j.\label{eq:JH2}
\end{align}
Furthermore,
\begin{align}
H(\mathsf{U}^{(1,d)})& \stackrel{(a)}{=} H(\mathsf{U}^{(1,d)},\mathsf{M}_{[\ell+1:d]})\nonumber\\
& \stackrel{(b)}{=} H(\mathsf{U}^{(1,d)}|\mathsf{M}_{[\ell+1:d]})+H(\mathsf{M}_{[\ell+1:d]})\nonumber\\
& \stackrel{(c)}{=} H(\mathsf{U}^{(1,d)}|\mathsf{M}_{[\ell+1:d]})+\sum_{j=\ell+1}^dB_j\nonumber\\
& \stackrel{(d)}{=} H(\mathsf{U}^{(1,d)},\mathsf{S}_{1\rightarrow[2:d-1]}|\mathsf{M}_{[\ell+1:d]})+\sum_{j=\ell+1}^dB_j\nonumber\\
& \stackrel{(e)}{=} H(\mathsf{U}^{(d-1)},\mathsf{W}_{d+1}|\mathsf{M}_{[\ell+1:d]})+\sum_{j=\ell+1}^dB_j\nonumber\\
& \geq H(\mathsf{U}^{(\ell)}|\mathsf{M}_{[\ell+1:d]})+\sum_{j=\ell+1}^dB_j\nonumber\\
& \stackrel{(f)}{=} H(\mathsf{U}^{(\ell)})+\sum_{j=\ell+1}^dB_j\label{eq:JH3}
\end{align}
where $(a)$ follows from the fact that $\mathsf{M}_{[\ell+1:d]}$ is a function of $\mathsf{W}_{[1:d]}$, which is is a function of $\mathsf{U}^{(1,d)}$ by Lemma~\ref{lemma1}; $(b)$ follows from the chain rule for entropy; $(c)$ follows from the fact that 
$H(\mathsf{M}_{[\ell+1:d]})=\sum_{j=\ell+1}^dB_j$; $(d)$ follows from the fact that $\mathsf{S}_{1\rightarrow[2:d-1]}$ is a function of $\mathsf{W}_1$ and hence a function of $\mathsf{U}^{(1,d)}$; $(e)$ follows from the fact that $H(\mathsf{U}^{(1,d)},\mathsf{S}_{1\rightarrow[2:d-1]}|\mathsf{M}_{[\ell+1:d]})=H(\mathsf{U}^{(d-1)},\mathsf{W}_{d+1}|\mathsf{M}_{[\ell+1:d]})$ due to the symmetrical code that we consider; and $(f)$ follows from the fact that $I(\mathsf{U}^{(\ell)};\mathsf{M}_{[\ell+1:d]})=0$ due to the repair secrecy requirement \eqref{eq:RepairSecrecy}.

Substituting \eqref{eq:JH2} and \eqref{eq:JH3} into \eqref{eq:JH1} gives:
\begin{align*}
& H(\mathsf{U}^{(1,\ell+1)})+\frac{T_{d,d,\ell+1}}{d-\ell}H(\mathsf{U}^{(\ell+1)})\\
& \geq \sum_{j=\ell+1}^{d-1}T_{d,j,\ell}^{-1}T_{d,d,j}B_j+\sum_{j=\ell+1}^dB_j+\left(1+\frac{T_{d,d,\ell+1}}{d-\ell}\right)H(\mathsf{U}^{(\ell)})\\
& = \sum_{j=\ell+1}^{d-1}T_{d,j,\ell}^{-1}(T_{d,d,j}+T_{d,j,\ell})B_j+B_d+\frac{T_{d,d,\ell}}{d-\ell}H(\mathsf{U}^{(\ell)})\\
& \stackrel{(a)}{=} T_{d,d,\ell}\sum_{j=\ell+1}^{d-1}T_{d,j,\ell}^{-1}B_j+B_d+\frac{T_{d,d,\ell}}{d-\ell}H(\mathsf{U}^{(\ell)})\\
& = T_{d,d,\ell}\sum_{j=\ell+1}^{d}T_{d,j,\ell}^{-1}B_j+\frac{T_{d,d,\ell}}{d-\ell}H(\mathsf{U}^{(\ell)})
\end{align*}
where $(a)$ follows from the fact that $T_{d,d,j}+T_{d,j,\ell}=T_{d,d,\ell}$. This completes the proof of the proposition. 
\end{IEEEproof}

We are now ready to prove the outer bounds \eqref{eq:B3} and \eqref{eq:B4}. To prove \eqref{eq:B3}, note that
\begin{align*}
\beta+\frac{1}{d-\ell}H(\mathsf{U}^{\ell}) & \stackrel{(a)}{\geq} \frac{1}{d-\ell}\left(H(\overline{\mathsf{S}}_{\rightarrow \ell+1})+H(\mathsf{U}^{(\ell)})\right)\\
& \stackrel{(b)}{\geq} \frac{1}{d-\ell}H(\mathsf{U}^{(\ell+1)})\\
& \stackrel{(c)}{\geq} \sum_{j=\ell+1}^{d}T_{d,j,\ell}^{-1}B_j+\frac{1}{d-\ell}H(\mathsf{U}^{(\ell)})
\end{align*}
where $(a)$ follows from the fact that $H(\overline{\mathsf{S}}_{\rightarrow \ell+1}) \leq (d-\ell)\beta$; $(b)$ follows from the union bound on entropy; and $(c)$ follows from \eqref{eq:prop1} of Proposition~\ref{prop1}. Cancelling $\frac{1}{d-\ell}H(\mathsf{U}^{\ell})$ from both sides of the inequality and normalizing both sides by $\sum_{t=\ell+1}^{d}B_t$ complete the proof of \eqref{eq:B3}.

To prove \eqref{eq:B4}, note that
\begin{align*}
\alpha&+(d(d-\ell)-\ell)\beta+(d+1)H(\mathsf{U}^{(\ell)})\\
& \stackrel{(a)}{\geq} H(\mathsf{W}_{d+1})+H(\mathsf{U}^{(\ell)})+(d(d-\ell)-\ell)\beta+dH(\mathsf{U}^{(\ell)})\\
& \stackrel{(b)}{=} H(\mathsf{W}_{d+1},\mathsf{S}_{d+1\rightarrow[1:\ell]})+H(\mathsf{U}^{(\ell)})+\\
& \hspace{20pt} (d(d-\ell)-\ell)\beta+dH(\mathsf{U}^{(\ell)})\\
& \stackrel{(c)}{\geq} H(\mathsf{W}_{d+1},\mathsf{U}^{(\ell)})+H(\mathsf{S}_{d+1\rightarrow[1:\ell]})+\\
& \hspace{20pt} (d(d-\ell)-\ell)\beta+dH(\mathsf{U}^{(\ell)})\\
& \stackrel{(d)}{\geq} H(\mathsf{W}_{d+1},\mathsf{U}^{(\ell)})+dH(\mathsf{U}^{(\ell+1)})\\
& \stackrel{(e)}{=} H(\mathsf{U}^{(1,\ell+1)},\mathsf{S}_{1\rightarrow[2:\ell+1]})+dH(\mathsf{U}^{(\ell+1)})\\
& \geq H(\mathsf{U}^{(1,\ell+1)})+dH(\mathsf{U}^{(\ell+1)})\\
& = H(\mathsf{U}^{(1,\ell+1)})+\frac{T_{d,d,\ell+1}}{d-\ell}H(\mathsf{U}^{(\ell+1)})+\\
& \hspace{20pt} \left(d-\frac{T_{d,d,\ell+1}}{d-\ell}\right)H(\mathsf{U}^{(\ell+1)})\\
& \stackrel{(f)}{\geq} T_{d,d,\ell}\left(\sum_{j=\ell+1}^{d}T_{d,j,\ell}^{-1}B_j+\frac{H(\mathsf{U}^{(\ell)})}{d-\ell}\right)+\\
& \hspace{20pt} \left(d-\frac{T_{d,d,\ell+1}}{d-\ell}\right)\left((d-\ell)\sum_{j=\ell+1}^{d}T_{d,j,\ell}^{-1}B_j+H(\mathsf{U}^{(\ell)})\right)\\
& = \left(T_{d,d,\ell}+d(d-\ell)-T_{d,d,\ell+1}\right)\sum_{j=\ell+1}^{d}T_{d,j,\ell}^{-1}B_j+\\
& \hspace{20pt} \left(\frac{T_{d,d,\ell}}{d-\ell}+d-\frac{T_{d,d,\ell+1}}{d-\ell}\right)H(\mathsf{U}^{(\ell)})\\
& \stackrel{(g)}{=} (d+1)(d-\ell)\sum_{j=\ell+1}^{d}T_{d,j,\ell}^{-1}B_j+(d+1)H(\mathsf{U}^{(\ell)})
\end{align*}
where $(a)$ follows from the fact that $H(\mathsf{W}_{d+1}) \leq \alpha$; $(b)$ follows from the fact that $\mathsf{S}_{d+1\rightarrow[1:\ell]}$ is a function of $\mathsf{W}_{d+1}$; $(c)$ follows from the fact that $H(\mathsf{W}_{d+1},\mathsf{S}_{d+1\rightarrow[1:\ell]})+H(\mathsf{U}^{(\ell)}) \geq H(\mathsf{W}_{d+1},\mathsf{U}^{(\ell)})+H(\mathsf{S}_{d+1\rightarrow[1:\ell]})$ due to the submodularity of the entropy function; $(d)$ follows from Proposition~\ref{prop2}; $(e)$ follows from the fact that $H(\mathsf{W}_{d+1},\mathsf{U}^{(\ell)})=H(\mathsf{U}^{(1,\ell+1)},\mathsf{S}_{1\rightarrow[2:\ell+1]})$ due to the symmetrical code that we consider; $(f)$ follows from \eqref{eq:prop1} of Proposition~\ref{prop1} and \eqref{eq:prop3} of Proposition~\ref{prop3}; and $(g)$ follows from the fact that $T_{d,d,\ell}-T_{d,d,\ell+1}=d-\ell$. Cancelling $(d+1)H(\mathsf{U}^{\ell})$ from both sides of the inequality and normalizing both sides by $\sum_{t=\ell+1}^{d}B_t$ complete the proof of \eqref{eq:B4}.

\section{Concluding remarks}\label{sec:Con}
This paper considered the problem of MDC-SR, which includes the problems of MDC-R and SRC as special cases. Two outer bounds were established, showing that separate coding can achieve the MBR point of the achievable normalized storage-capacity repair-bandwidth tradeoff regions for the general MDC-SR problem. When specialized to the SRC problem, it was shown that the SRK point \cite{Shah-Globecom11} is the MBR point of the achievable normalized storage-capacity repair-bandwidth tradeoff regions for the general SRC problem. The core of the new converse results is an exchange lemma, which we established by using Han's subset inequality \cite{Han-IC78}. The exchange lemma only relies on the functional dependencies for the repair processes and might be useful for solving some other related problems as well. 

Note that separate encoding can also achieve the MSR point of the achievable normalized storage-capacity repair-bandwidth tradeoff regions for the general MDC-R problem \cite{Shao-CISS16}. We suspect that this also generalizes to the MDC-SR problem. To prove such this result, however, we shall need new converse results as well as new code constructions for the general SRC problem, both of which are currently under our investigations.

\appendix[Proof of the Exchange Lemma]

\begin{figure}[t!]
\centering
\includegraphics[width=0.5\linewidth,draft=false]{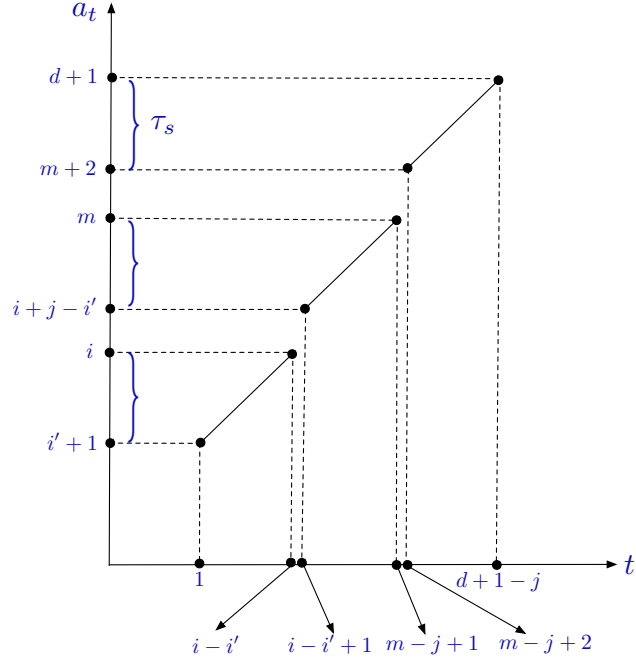}
\caption{$a_t$ as a function of $t$. The sets $(\tau_q: q\in[0:s])$ form a partition of the set $[i'+1:i]\cup [i+j-i':m]\cup [m+2:d+1]$.}
\vspace{0.3cm}
\label{fig3}
\end{figure}

Fix $m\in[1:d-1]$, $i\in[0:m-1]$, $i'\in[0:i]$, and $j\in[i'+1:m-i+i'+1]$. Let us first note that if $j=m+1$, we must have $i'=i$, and in this case the inequality \eqref{eq:exchange} holds trivially with an equality. Therefore, for the remaining proof we shall assume that $j \leq m$. Now that $d+1-j > d-m$, we may write $d+1-j=s(d-m)+r$ for some integer $s\geq 1$ and $r\in[1:d-m]$. Furthermore, let
$$ a_t:=\left\{
\begin{array}{lcl}
t+i' ,     &      & {t     \in   [1:i-i']   }\\
t+j-1,   &      & {t \in [i-i'+1: m-j+1]}\\
t+j ,      &      & {t \in [m-j+2: d+1-j]}.
\end{array} \right. $$
As illustrated in Figure~\ref{fig3}, $a_t$ is monotonically increasing with $t$. Finally, let $\tau_0:=\{a_t: t\in [1: r]\}$ and
\begin{align*}
\tau_q:=\{a_t: t\in [r+1+(q-1)(d-m):r+q(d-m)]\}
\end{align*} 
for any $q\in[1:s]$. It is straightforward to verify that:

\begin{itemize}
\item $\tau_q  \cap  \tau_{q'} =\emptyset$ for any $q\neq q'$;
\item $\bigcup_{q=0}^{s-1}\tau_q =[i'+1:i]\cup [i+j-i':m]$;
\item $\tau_s=[m+2:d+1]$.
\end{itemize}

Consider a symmetrical $(n=d+1,d,N_1,\ldots,N_d,T,S)$ code that satisfies the node regeneration requirement \eqref{eq:NodeRegen}. Let us show by induction that for any $p\in[1:s]$, we have
\begin{align}
&pH(\mathsf{U}^{(i,m)}|\mathsf{M}^{(m)})+H(\mathsf{U}^{(i',j)}|\mathsf{M}^{(m)})\nonumber\\
& \geq pH(\mathsf{U}^{(i,m+1)}|\mathsf{M}^{(m)})+\nonumber\\
& \hspace{20pt} H(\mathsf{W}_{[1:i']},\mathsf{S}_{\rightarrow [i+1:i+j-i'-1]},\mathsf{S}_{\bigcup_{q=0}^{s-p}\tau_q\rightarrow m+1}|\mathsf{M}^{(m)}).\label{eq:QQ1}
\end{align}

To prove the base case of $p=1$, first note that
\begin{align*}
H&(\mathsf{U}^{(i,m)}|\mathsf{M}^{(m)})\\
&\stackrel{(a)}{=} H(\mathsf{U}^{(i,m)},\mathsf{W}_{[i+1,m]},\underline{\mathsf{S}}_{\rightarrow [i+1:m]}|\mathsf{M}^{(m)})\\
&= H(\mathsf{W}_{[1:m]},\mathsf{S}_{\rightarrow [i+1:m]}|\mathsf{M}^{(m)})\\
&\stackrel{(b)}{=} H(\mathsf{W}_{[1:m]},\mathsf{S}_{\rightarrow [i+1:m]},\mathsf{S}_{[1:m]\rightarrow m+1}|\mathsf{M}^{(m)}))\\
& \geq H(\mathsf{W}_{[1:i]},\mathsf{S}_{\rightarrow [i+1:m]},\mathsf{S}_{[1:m]\rightarrow m+1}|\mathsf{M}^{(m)}))
\end{align*}
where $(a)$ follows from the fact that $(\mathsf{W}_{[i+1,m]},\underline{\mathsf{S}}_{\rightarrow [i+1:m]})$ is a function of $\mathsf{U}^{(i,m)}$ by Lemma~\ref{lemma1}, and $(b)$ follows from the fact that $\mathsf{S}_{[1:m]\rightarrow m+1}$ is a function of $\mathsf{W}_{[1:m]}$. Furthermore,
\begin{align*}
H&(\mathsf{U}^{(i',j)}|\mathsf{M}^{(m)})\\
&\stackrel{(a)}{=} H(\mathsf{U}^{(i',j)},\underline{\mathsf{S}}_{\rightarrow [i'+1:j]}|\mathsf{M}^{(m)})\\
&= H(\mathsf{W}_{[1:i']},\mathsf{S}_{\rightarrow [i'+1:j]}|\mathsf{M}^{(m)})\\
&\stackrel{(b)}{=} H(\mathsf{W}_{[1:i']},\mathsf{S}_{\rightarrow [i+1:i+j-i']}|\mathsf{M}^{(m)})\\
&= H(\mathsf{W}_{[1:i']},\mathsf{S}_{\rightarrow [i+1:i+j-i'-1]}, \mathsf{S}_{\rightarrow i+j-i'}|\mathsf{M}^{(m)})\\
&\geq H(\mathsf{W}_{[1:i']},\mathsf{S}_{\rightarrow [i+1:i+j-i'-1]}, \mathsf{S}_{[i'+1:i]\rightarrow i+j-i'}, \\
& \hspace{20pt}\mathsf{S}_{[i+j-i'+1:d+1]\rightarrow i+j-i'}|\mathsf{M}^{(m)})\\
&\stackrel{(c)}{=} H(\mathsf{W}_{[1:i']},\mathsf{S}_{\rightarrow [i+1:i+j-i'-1]}, \mathsf{S}_{[i'+1:i]\rightarrow m+1},\\
& \hspace{20pt}\mathsf{S}_{[i+j-i':m]\rightarrow m+1}, \mathsf{S}_{[m+2:d+1]\rightarrow m+1}|\mathsf{M}^{(m)})
\end{align*}
where $(a)$ follows from the fact that $\underline{\mathsf{S}}_{\rightarrow [i'+1:j]}$ is a function of $\mathsf{U}^{(i',j)}$ by Lemma~\ref{lemma1}, and $(b)$ and $(c)$ follow from the symmetrical code that we consider. It follows that
\begin{align*}
&H(\mathsf{U}^{(i,m)}|\mathsf{M}^{(m)})+H(\mathsf{U}^{i',j}|\mathsf{M}^{(m)})\\
&\geq H(\mathsf{W}_{[1:i]},\mathsf{S}_{\rightarrow [i+1:m]},\mathsf{S}_{[1:m]\rightarrow m+1}|\mathsf{M}^{(m)}))+\\
& \hspace{20pt} H(\mathsf{W}_{[1:i']},\mathsf{S}_{\rightarrow [i+1:i+j-i'-1]},\mathsf{S}_{[i'+1:i]\rightarrow m+1},\\
& \hspace{20pt} \mathsf{S}_{[i+j-i':m]\rightarrow m+1}, \mathsf{S}_{[m+2:d+1]\rightarrow m+1}|\mathsf{M}^{(m)})\\
& \stackrel{(a)}{\geq} H(\mathsf{W}_{[1:i]},\mathsf{S}_{\rightarrow [i+1:m]},\mathsf{S}_{[1:m]\rightarrow m+1},\\
& \hspace{20pt}\mathsf{S}_{[m+2:d+1]\rightarrow m+1}|\mathsf{M}^{(m)})+H(\mathsf{W}_{[1:i']},\mathsf{S}_{\rightarrow [i+1:i+j-i'-1]},\\
& \hspace{20pt} \mathsf{S}_{[i'+1:i]\rightarrow m+1},\mathsf{S}_{[i+j-i':m]\rightarrow m+1}|\mathsf{M}^{(m)})\nonumber\\
& = H(\mathsf{U}^{(i,m+1)},\underline{\mathsf{S}}_{\rightarrow[i+1:m+1]}|\mathsf{M}^{(m)})+\\
& \hspace{20pt} H(\mathsf{W}_{[1:i']},\mathsf{S}_{\rightarrow [i+1:i+j-i'-1]},\mathsf{S}_{\bigcup_{q=0}^{s-1}\tau_q\rightarrow m+1}|\mathsf{M}^{(m)})\\
& \geq H(\mathsf{U}^{(i,m+1)}|\mathsf{M}^{(m)})+\\
& \hspace{20pt} H(\mathsf{W}_{[1:i']},\mathsf{S}_{\rightarrow [i+1:i+j-i'-1]},\mathsf{S}_{\bigcup_{q=0}^{s-1}\tau_q\rightarrow m+1}|\mathsf{M}^{(m)})
\end{align*}
where $(a)$ follows from the submodularity of the entropy function. This completes the proof of the base case of $p=1$.

Assume that \eqref{eq:QQ1} holds for some $p\in[1:s-1]$. We have
\begin{align}
&(p+1)H(\mathsf{U}^{(i,m)}|\mathsf{M}^{(m)})+H(\mathsf{U}^{(i',j)}|\mathsf{M}^{(m)})\nonumber\\
& = H(\mathsf{U}^{(i,m)}|\mathsf{M}^{(m)})+\left(pH(\mathsf{U}^{(i,m)}|\mathsf{M}^{(m)})+H(\mathsf{U}^{(i',j)}|\mathsf{M}^{(m)})\right)\nonumber\\
& \geq H(\mathsf{U}^{(i,m)}|\mathsf{M}^{(m)})+pH(\mathsf{U}^{(i,m+1)}|\mathsf{M}^{(m)})+\nonumber\\
& \hspace{20pt} H(\mathsf{W}_{[1:i']},\mathsf{S}_{\rightarrow [i+1:i+j-i'-1]},\mathsf{S}_{\bigcup_{q=0}^{s-p}\tau_q\rightarrow m+1}|\mathsf{M}^{(m)}).\label{eq:QQ2}
\end{align}
Note that both $\underline{\mathsf{S}}_{\rightarrow[i+1,i+j-i'-1]}$ and $\mathsf{S}_{\bigcup_{q=0}^{s-(p+1)}\tau_q\rightarrow m+1}$ are functions of $\mathsf{W}_{[1:m]}$, which is in turn a function of $\mathsf{U}^{(i,m)}$ by Lemma~\ref{lemma1}. We thus have
\begin{align*}
H&(\mathsf{U}^{(i,m)}|\mathsf{M}^{(m)})\\
&=H(\mathsf{U}^{(i,m)},\underline{\mathsf{S}}_{\rightarrow[i+1,i+j-i'-1]},\mathsf{S}_{\bigcup_{q=0}^{s-(p+1)}\tau_q\rightarrow m+1}|\mathsf{M}^{(m)}).
\end{align*}
Furthermore, by the symmetrical code that we consider we have
\begin{align*}
H&(\mathsf{W}_{[1:i']},\mathsf{S}_{\rightarrow [i+1:i+j-i'-1]},\mathsf{S}_{\bigcup_{q=0}^{s-p}\tau_q\rightarrow m+1}|\mathsf{M}^{(m)})\\
& = H(\mathsf{W}_{[1:i']},\mathsf{S}_{\rightarrow [i+1:i+j-i'-1]},\\
& \hspace{20pt} \mathsf{S}_{\bigcup_{q=0}^{s-(p+1)}\tau_q\rightarrow m+1},\mathsf{S}_{[m+2:d+1]\rightarrow m+1}|\mathsf{M}^{(m)}).
\end{align*}
It follows that 
\begin{align}
&H(\mathsf{U}^{(i,m)}|\mathsf{M}^{(m)})+\nonumber\\
& \hspace{20pt} H(\mathsf{W}_{[1:i']},\mathsf{S}_{\rightarrow [i+1:i+j-i'-1]},\mathsf{S}_{\bigcup_{q=0}^{s-p}\tau_q\rightarrow m+1}|\mathsf{M}^{(m)})\nonumber\\
& = H(\mathsf{U}^{(i,m)},\underline{\mathsf{S}}_{\rightarrow[i+1,i+j-i'-1]},\mathsf{S}_{\bigcup_{q=0}^{s-(p+1)}\tau_q\rightarrow m+1}|\mathsf{M}^{(m)})+\nonumber\\
& \hspace{20pt} H(\mathsf{W}_{[1:i']},\mathsf{S}_{\rightarrow [i+1:i+j-i'-1]},\nonumber\\
& \hspace{20pt} \mathsf{S}_{\bigcup_{q=0}^{s-(p+1)}\tau_q\rightarrow m+1},\mathsf{S}_{[m+2:d+1]\rightarrow m+1}|\mathsf{M}^{(m)})\nonumber\\
& \stackrel{(a)}{\geq} H(\mathsf{U}^{(i,m)},\underline{\mathsf{S}}_{\rightarrow[i+1,i+j-i'-1]},\nonumber\\
& \hspace{20pt} \mathsf{S}_{\bigcup_{q=0}^{s-(p+1)}\tau_q\rightarrow m+1},\mathsf{S}_{[m+2:d+1]\rightarrow m+1}|\mathsf{M}^{(m)})+\nonumber\\
& \hspace{20pt} H(\mathsf{W}_{[1:i']},\mathsf{S}_{\rightarrow [i+1:i+j-i'-1]},\mathsf{S}_{\bigcup_{q=0}^{s-(p+1)}\tau_q\rightarrow m+1}|\mathsf{M}^{(m)})\nonumber\\
& \geq H(\mathsf{U}^{(i,m+1)}|\mathsf{M}^{(m)})+\nonumber\\
& \hspace{20pt} H(\mathsf{W}_{[1:i']},\mathsf{S}_{\rightarrow [i+1:i+j-i'-1]},\mathsf{S}_{\bigcup_{q=0}^{s-(p+1)}\tau_q\rightarrow m+1}|\mathsf{M}^{(m)})\label{eq:QQ3}
\end{align}
where $(a)$ follows from the submodularity of the entropy function. Substituting \eqref{eq:QQ3} into \eqref{eq:QQ2} gives
\begin{align*}
&(p+1)H(\mathsf{U}^{(i,m)}|\mathsf{M}^{(m)})+H(\mathsf{U}^{(i',j)}|\mathsf{M}^{(m)})\\
& \geq (p+1)H(\mathsf{U}^{(i,m+1)}|\mathsf{M}^{(m)})+\nonumber\\
& \hspace{20pt} H(\mathsf{W}_{[1:i']},\mathsf{S}_{\rightarrow [i+1:i+j-i'-1]},\mathsf{S}_{\bigcup_{q=0}^{s-(p+1)}\tau_q\rightarrow m+1}|\mathsf{M}^{(m)})
\end{align*}
which completes the induction step and hence the proof of \eqref{eq:QQ1}.

Setting $p=s$ in \eqref{eq:QQ1}, we have
\begin{align}
&sH(\mathsf{U}^{(i,m)}|\mathsf{M}^{(m)})+H(\mathsf{U}^{(i',j)}|\mathsf{M}^{(m)})\nonumber\\
& \geq sH(\mathsf{U}^{(i,m+1)}|\mathsf{M}^{(m)})+\nonumber\\
& \hspace{20pt} H(\mathsf{W}_{[1:i']},\mathsf{S}_{\rightarrow [i+1:i+j-i'-1]},\mathsf{S}_{\tau_0\rightarrow m+1}|\mathsf{M}^{(m)})\nonumber\\
& = sH(\mathsf{U}^{(i,m+1)}|\mathsf{M}^{(m)})+H(\mathsf{W}_{[1:i']},\mathsf{S}_{\rightarrow [i+1:i+j-i'-1]}|\mathsf{M}^{(m)})+\nonumber\\
& \hspace{20pt} H(\mathsf{S}_{\tau_0\rightarrow m+1}|\mathsf{W}_{[1:i']},\mathsf{S}_{\rightarrow [i+1:i+j-i'-1]},\mathsf{M}^{(m)}).\label{eq:QQ4}
\end{align}
By the symmetrical codes that we consider, we have
\begin{align}
H&(\mathsf{W}_{[1:i']},\mathsf{S}_{\rightarrow [i+1:i+j-i'-1]}|\mathsf{M}^{(m)})\nonumber\\
&=H(\mathsf{W}_{[1:i']},\mathsf{S}_{\rightarrow [i'+1:j-1]}|\mathsf{M}^{(m)})\nonumber\\
&= H(\mathsf{U}^{(i',j-1)},\underline{\mathsf{S}}_{\rightarrow [i'+1:j-1]}|\mathsf{M}^{(m)})\nonumber\\
&\geq H(\mathsf{U}^{(i',j-1)}|\mathsf{M}^{(m)})\label{eq:QQ5}
\end{align}
and
\begin{align*}
&H(\mathsf{S}_{\tau_0\rightarrow m+1}|\mathsf{W}_{[1:i']},\mathsf{S}_{\rightarrow [i+1:i+j-i'-1]},\mathsf{M}^{(m)})\\
&=H(\mathsf{S}_{\tau\rightarrow m+1}|\mathsf{W}_{[1:i']},\mathsf{S}_{\rightarrow [i+1:i+j-i'-1]},\mathsf{M}^{(m)}
\end{align*}
for {\em any} subset $\tau \subseteq [m+2:d+1]$ such that $|\tau|=r$. By Han's subset inequality \cite{Han-IC78}, we have
\begin{align}
H&(\mathsf{S}_{\tau_0\rightarrow m+1}|\mathsf{W}_{[1:i']},\mathsf{S}_{\rightarrow [i+1:i+j-i'-1]},\mathsf{M}^{(m)})\nonumber\\
& \geq \frac{r}{d-m}H(\mathsf{S}_{[m+2:d+1]\rightarrow m+1}|\mathsf{W}_{[1:i']},\nonumber\\
& \hspace{20pt}\mathsf{S}_{\rightarrow [i+1:i+j-i'-1]},\mathsf{M}^{(m)})\nonumber\\
& \geq \frac{r}{d-m}H(\mathsf{S}_{[m+2:d+1]\rightarrow m+1}|\mathsf{W}_{[1:i']},\nonumber\\
&\hspace{20pt} \mathsf{S}_{\rightarrow [i+1:i+j-i'-1]},\mathsf{U}^{(i,m)},\mathsf{M}^{(m)})\nonumber\\
& \stackrel{(a)}{=} \frac{r}{d-m}H(\mathsf{S}_{[m+2:d+1]\rightarrow m+1}|\mathsf{U}^{(i,m)},\mathsf{M}^{(m)})\nonumber\\
& = \frac{r}{d-m}\left(H(\mathsf{S}_{[m+2:d+1]\rightarrow m+1},\mathsf{U}^{(i,m)}|\mathsf{M}^{(m)})-\right.\nonumber\\
& \hspace{20pt} \left.H(\mathsf{U}^{(i,m)}|\mathsf{M}^{(m)})\right)\nonumber\\
& = \frac{r}{d-m}\left(H(\mathsf{U}^{(i,m+1)}|\mathsf{M}^{(m)})-H(\mathsf{U}^{(i,m)}|\mathsf{M}^{(m)})\right)\label{eq:QQ6}
\end{align}
where $(a)$ follows from the fact that $(\mathsf{W}_{[1:i']},\underline{\mathsf{S}}_{\rightarrow [i+1:i+j-i'-1]})$ is a function of $\mathsf{U}^{(i,m)}$ by Lemma~\ref{lemma1}. Substituting \eqref{eq:QQ5} and \eqref{eq:QQ6} into \eqref{eq:QQ4} gives:
\begin{align*}
&\left(s+\frac{r}{d-m}\right)H(\mathsf{U}^{(i,m)}|\mathsf{M}^{(m)})+H(\mathsf{U}^{(i',j)}|\mathsf{M}^{(m)})\\
& \geq \left(s+\frac{r}{d-m}\right)H(\mathsf{U}^{(i,m+1)}|\mathsf{M}^{(m)})+H(\mathsf{U}^{(i',j-1)}|\mathsf{M}^{(m)})
\end{align*}
which is equivalent to \eqref{eq:exchange} by noting that
\begin{align*}
s+\frac{r}{d-m}=\frac{s(d-m)+r}{d-m}=\frac{d+1-j}{d-m}.
\end{align*}
This completes the proof of the exchange lemma.

\bibliographystyle{ieeetr}

\end{document}